\documentclass[12pt,preprint,natbib]{aastex}
\input epsf
\def\s2n{S^{\prime}/N}

\bibpunct{(}{)}{;}{a}{}{,}

\begin{document}
\title{The ``Mysterious'' Origin of Brown Dwarfs}

\author{Paolo Padoan\footnote{ppadoan@ucsd.edu},
\affil{Department of Physics, University of California, San Diego, CASS/UCSD 0424, 9500 Gilman Drive, La Jolla, CA 92093-0424} 
\AA ke Nordlund\footnote{aake@astro.ku.dk}}
\affil{Astronomical Observatory / NBIfAFG, Juliane Maries Vej 30, DK-2100, Copenhagen, Denmark}

\begin{abstract}

Hundreds of brown dwarfs (BDs) have been discovered in the last few years
in stellar clusters and among field stars. BDs are almost as numerous
as hydrogen burning stars and so a theory of star formation should
also explain their origin. The ``mystery'' of the origin of BDs
is that their mass is two orders of magnitude smaller than the average
Jeans' mass in star--forming clouds, and yet they are so common.
In this work we investigate the possibility that gravitationally unstable
protostellar cores of BD mass are formed directly by 
the process of turbulent fragmentation. 
Supersonic turbulence in molecular clouds generates a complex density
field with a very large density contrast. As a result, a fraction of 
BD mass cores formed by the turbulent flow are dense enough to be 
gravitationally unstable. We find that with density, temperature and 
rms Mach number typical of cluster--forming regions, turbulent fragmentation 
can account for the observed BD abundance.

\end{abstract}


\section{Introduction}

A large number of Brown Dwarfs (BDs) have been discovered in the last
few years, both within stellar clusters
\citep[e.g.][]{Bouvier+98,Martin+98,Zapatero-Osorio+2000,%
Luhman+2000,Hillenbrand+Carpenter2000,Najita+2000,Comeron+2000,%
Luhman2000,Moraux+2001,Bejar+2001,Martin+2001,Lodieu+2002}
and among field stars
\citep[e.g.][]{Kirkpatrick+99,Kirkpatrick+2000,Chabrier2002}.
It is now well established that BDs do not hide a significant amount
of baryonic dark matter, at least in our galaxy
\citep[e.g.][]{Najita+2000,Bejar+2001,Chabrier2002}.
The stellar initial mass function (IMF) is flat or decreasing 
toward sub--stellar masses.

Even if the total mass
of BDs is not dynamically important, their abundance relative to hydrogen
burning stars is so large that their existence cannot be overlooked in
the context of a theory of star formation. According to cluster and field
IMFs extended to sub--stellar masses, there are almost as
many BDs as regular stars \citep[e.g.][]{Bejar+2001,Chabrier2002,Chabrier2003}.
BDs may be relevant also for understanding the formation of planets,
because their mass is intermediate between that of hydrogen burning stars
($M>0.07$--$0.08$~M$_{\odot}$) and that of planets
($M<0.011$--$0.013$~M$_{\odot}$, using the deuterium burning limit to
separate planets from BDs).

Surprisingly, very little theoretical research has addressed the problem
of the formation of BDs. Most theoretical research on BDs is concerned
with modeling the structure and evolution of sub--stellar objects and their
atmospheres, in order to derive their observational properties
\citep[e.g.][]{Chabrier+2000,Chabrier+Baraffe2000,Baraffe+2002}

The classical Jeans' mass \citep{Jeans02} at the mean density and temperature of a typical
star--forming cloud is several solar masses.
Based on the usual assumption that the Jeans' mass is an
approximate estimate of the lower limit to the stellar mass
\citep{Larson92,Elmegreen99}, the formation of BDs is an unsolved problem.

Recent observational results have shown that the mass distribution of
prestellar condensations is indistinguishable from the stellar IMF
\citep{Motte+98,Testi+Sargent98,Onishi+99,Johnstone+2000a,Johnstone+2000b,%
Motte+2001,Johnstone+2001,Onishi+2002}, both in the functional shape and in the
range of masses, including BD mass cores \citep{Walsh+04}. 
The problem of the formation of BDs in molecular clouds
with a Jeans' mass much larger than the mass of BDs is thus unlikely to
be solved by relying on a significant mass difference between a single
collapsing core and its final star.

\cite{Elmegreen99} stressed the importance of BDs for testing star
formation theories. He proposed that BDs are more abundant in ultra-cold
regions in the inner disk of M31 or in spiral--arm dust lanes than in
normal star--forming clouds. He assumed that the smallest stellar mass
is of the order of the thermal Jeans' mass and therefore the BD abundance
should increase with decreasing gas temperature or increasing pressure.
His argument is consistent with the present work, where we focus on 
more typical regions of star formation.

\cite{Reipurth+Clarke2001} proposed that BDs are the leftover
of a prematurely interrupted accretion process. Their model assumes that
stars are born as multiple systems of small ``embryos'' sharing a
common reservoir of accreting gas. The accretion process of one of them
is ``aborted'' when it is ejected by gravitational interaction with a pair
of companions. One problem with this model is that protostellar disks are 
frequently found around BDs \citep{Natta+Testi2001,Natta+2002,Liu2003,Jayawardhana+2003,Jayawardhana2003a,Klein+2003,Martin+2004,LopezMarti+2004,Mohanty+2004}.  
Other problems posed by observational data are discussed in \cite{Briceno+2002}.

\cite{Bate+2002} have interpreted the results of
a numerical simulation as evidence in favor of the \cite{Reipurth+Clarke2001}
model. They simulated the collapse of a 50~M$_{\odot}$ isothermal gas
cloud, with an initial random velocity field and uniform density, using
a smooth particle hydrodynamic (SPH) code. They found that BDs are formed 
mainly as members of multiple systems by the gravitational fragmentation of a common
protostellar disk. In their simulation, the initial velocity field is not 
obtained as the solution of the fluid equations self--consistently with 
the density field. It is instead generated artificially by imposing a power spectrum
consistent with Larson's velocity-size correlation \citep{Larson79,Larson81}, 
while the initial density is uniform. As the turbulence is not externally 
driven, it decays while the cloud collapses and there is no time for the flow 
to rearrange as a realistic supersonic turbulent flow independent of the initial
conditions. Another problem is the absence of a magnetic field, which 
affects significantly the fragmentation process as the magnetic field
may modify the shock jump conditions and the angular momentum transport. 

Given the numerical limitations in the simulation by \cite{Bate+2002}, it is 
possible that turbulent fragmentation has been overlooked as the origin of 
collapsing cores of BD mass. In the present work we investigate the 
possibility that BDs are the direct consequence of turbulent 
fragmentation \citep{Padoan+Nordlund99MHD,%
Padoan+2001cores,Padoan+Nordlund2002IMF,Nordlund+Padoan2002Paris},
in the sense that they are assembled by the turbulent flow as 
gravitationally unstable objects. We first compute the upper limit to the 
BD abundance in \S~2, based on the probability density function (PDF) of 
gas density in supersonic turbulence. In \S~3 we compute the actual BD abundance,
according to our analytical model of the stellar IMF, and in \S~4 we
show that numerical simulations of supersonic magneto--hydrodynamic
(MHD) turbulence provide support for our analytical model. Results are
summarized and conclusions drawn in \S~5.

\section{An Upper Limit to the BD Abundance}

The gas density and velocity fields in star--forming clouds are highly
non--linear due to the presence of supersonic turbulence. The kinetic
energy of turbulence is typically 100 times larger than the gas thermal
energy on the scale of a few pc (the typical rms Mach number is of the
order of 10) and the gas is roughly isothermal, so that very large
compressions due to a complex network of interacting shocks cannot be
avoided. Under such conditions the concept of gravitational instability,
based on a comparison between gravitational and thermal energies alone
in a system with mild perturbations, does not apply.
Dense cores of any size can be formed in the turbulent flow, independent of the Jeans' mass.
Those cores that are massive and dense enough (larger than their own
Jeans' mass) collapse into protostars, while smaller subcritical
ones re-expand into the turbulent flow. This is a process that we call
{\it turbulent fragmentation}, to stress the point that stars and BDs,
formed in supersonically turbulent clouds, are not primarily the 
result of gravitational fragmentation.

Nevertheless, the idea of a critical mass for gravitational collapse
is often applied to star forming clouds. The critical mass is defined
as the Jeans' or the Bonnor--Ebert mass \citep{Bonnor56,McCrea57} 
computed with the average density, temperature and pressure. The effect 
of the kinetic energy of turbulence is modeled as an external pressure, 
$P_t\sim\rho \sigma_{\rm v}^2$,
where $\rho$ is the gas density and $\sigma_{\rm v}$ is the rms velocity of
the turbulence. \cite{Elmegreen99}, for example, has proposed that the
minimum stellar mass is of the order of this Bonnor--Ebert mass.
The minimum stellar mass in typical star--forming clouds is then 
several ~M$_{\odot}$, which contradicts the relatively large
abundance of BDs.

Dense cores formed by the turbulent flow need to be larger than
their thermal critical mass to collapse (neglecting the magnetic field). 
Therefore, a necessary condition for
the formation of BDs by supersonic turbulence is the existence of
a finite mass fraction, in the turbulent flow, with
density at least as high as the critical one for the collapse
of a BD mass core. As an estimate of the critical mass we use the 
Bonnor--Ebert mass, instead of the classical Jeans' mass, because 
the cores assembled by the turbulent flow are non--linear density
enhancements bounded by the shock ram pressure, rather than linear
density perturbations as in Jeans' assumption. The mass, $m_{\rm BE}$, 
of the critical Bonnor--Ebert isothermal sphere is \citep{Bonnor56}: 
\begin{equation}
m_{\rm BE}=3.3\, {\rm M}_{\odot}\left(\frac{T}{10\, K}\right)^{3/2}
        \left(\frac{n}{10^3\, cm^{-3}}\right)^{-1/2},
\label{mj0}
\end{equation}
We have verified numerically that this expression is a good estimate of the 
critical mass, both in simple geometries and in turbulence simulations 
including selfgravity.

We want to estimate the mass fraction with density at least as high 
as the critical one for the collapse of a BD mass core. Using the
value $m_{\rm BD}=0.075$~M$_{\odot}$ for the largest BD mass,
the condition $m_{\rm BE}<m_{\rm BD}=0.075$~M$_{\odot}$ corresponds to
$n > n_{\rm BD}=1.8\times 10^6$~cm$^{-3}$, using equation (\ref{mj0}). 
The fraction of the total mass of the system that can form BDs is given 
by the probability that $n > n_{BD}$. This is $\int_{n_{\rm BD}}^{\infty} p(n)\,dn$, 
where $p(n)$ is the probability density function (PDF) of the gas density $n$.
This probability can be interpreted as a fractional volume and therefore 
the mass fraction available for the formation of BDs is:
\begin{equation}
f_{\rm BD}={\int_{n_{\rm BD}}^{\infty} n\,p(n)\,dn \over \int_{0}^{\infty} n\,p(n)\,dn}
\label{fBD}
\end{equation}
It is important to stress that $f_{\rm BD}$ is {\it not} the BD mass
fraction, but only the mass fraction of the gas that {\it could}
end up in BDs, or the upper limit for the BD mass fraction
(assuming BDs are due only to turbulent fragmentation).
Most likely the star formation process is rather inefficient
for BDs as it is for hydrogen burning stars, and so the actual
fractional mass in BDs may be significantly smaller than $f_{BD}$.

One of the most important universal properties of turbulent fragmentation
is that the PDF of gas density is Log--Normal for an isothermal gas, so
that
\begin{equation}
p(n)dn=\frac{1/n}{(2\pi\sigma^{2})^{1/2}}exp\left[-\frac{1}{2}
\left(\frac{\ln n-\overline{\ln n}}{\sigma}\right)^{2}\right]dn,
\label{pdf1}
\end{equation}
for the case of $\langle n\rangle=1$. The average value of the logarithm
of density, $\overline{\ln n}$, is determined by the standard deviation
$\sigma$ of the logarithm of density (a property of the Log--Normal,
here used again for the case $\langle n\rangle=1$):
\begin{equation}
\overline{\ln n}=-\frac{\sigma^{2}}{2} ,
\label{pdf3}
\end{equation}
and the standard deviation of the logarithm of density, $\sigma$,
is a function of the rms sonic Mach number of the flow, $M_{\rm S}$:
\begin{equation}
\sigma^{2}=\ln(1+b^2 M_{\rm S}^2)
\label{pdf4}
\end{equation}
or, equivalently, the standard deviation of the linear density is:
\begin{equation}
\sigma_{\rho}=b M_{\rm S}
\label{pdf5}
\end{equation}
where $b\approx0.5$ \citep{Nordlund+Padoan_Puebla98,Ostriker+99}.
This result is very useful because the rms Mach
number of the turbulence in molecular clouds is easily estimated
through the spectral line width of molecular transitions and by
estimating the kinetic temperature. The three dimensional PDF
of gas density is then fully determined by the value of the rms
Mach number.

A contour plot of $f_{\rm BD}$, computed from (\ref{fBD}) and with the
Log--Normal PDF (\ref{pdf1}), on the plane $\langle n \rangle$--$M_{\rm S}$,
is shown in Figure~\ref{fig1}. The dotted line corresponds
to values of rms Mach number and average gas density of typical Larson
relations \citep{Larson81,Brunt03,Heyer+Brunt04}. Figure~\ref{fig1} shows that 
in regions following the average Larson relations, approximately
1\% of the total mass is available for the formation of BDs (dotted
line in Figure~\ref{fig1}). If most of the available 1\% of the total 
mass were turned into BDs, the number of BDs would be comparable to the 
number of hydrogen burning stars even if as much as 10\% of the total
mass was turned into stars. However, it is possible that the formation
of BDs has a typical efficiency of only a few percent, similar to that 
of hydrogen burning stars. In that case their relative abundance in 
molecular clouds following the average Larson relations is expected
to be rather low. 

For a given size and velocity dispersion, cluster--forming regions 
are a few times denser than clouds following the average density--size 
Larson relation. In cluster--forming regions, the mass available for 
the formation of BDs can be very large, $f_{\rm BD}\sim 0.1$. 
As an example, the stellar mass density in the central $5\times 5$~arcmin 
($\approx 0.35\times0.35$~pc) of the young cluster IC 348 \citep{Luhman+2003} 
corresponds approximately to $2\times 10^4$~cm$^{-3}$. 
Because the star formation efficiency is likely to be less than unity,
the initial gas density in that region must have been even larger. 
If we assume a gas density of $5\times 10^4$~cm$^{-3}$ (corresponding to a 
star formation efficiency of 40\%) and a velocity dispersion taken from 
the average velocity--size Larson relation at a size equal to 0.35~pc, 
we get $f_{\rm BD}\approx 0.1$, as shown by the square in Figure~\ref{fig1}.    
Other cluster--forming regions, such as the central region of the Trapezium 
cluster \citep{Hillenbrand+Carpenter2000,Luhman+2000}, show similar
stellar mass densities as IC 348. With such a large value of $f_{\rm BD}$, 
cluster--forming regions may in principle form as many BDs as hydrogen 
burning stars, even if the efficiency of BD formation (from the gas with 
density larger than $n_{\rm BD}$) is as low as the star formation 
efficiency for hydrogen burning stars. 

The cross in the middle of Figure~\ref{fig1} shows the values of 
$\langle n \rangle$ and $M_{\rm S}$ corresponding to the initial conditions
in the simulation by \cite{Bate+2002}. They simulate a cloud with a total
mass of 50~M$_{\odot}$, a diameter of 0.375~pc, a mean molecular weight
of the gas of 2.46, a temperature of 10~K and a kinetic energy of the
turbulence equal to the cloud gravitational potential energy. From
these initial conditions we obtain $\langle n \rangle=3.35\times 10^4$~cm$^{-3}$
and rms sonic Mach number of the turbulence $M_{\rm S}=6.5$. The contour
plot indicates that 5\% of the total mass is in this case available for the formation
of BDs, sufficient to generate as many BDs as hydrogen burning stars, as discussed
above. This suggests that under the conditions assumed in that simulation BDs could
originate in large abundance as the result of the turbulent fragmentation.

\nocite{Larson81,Falgarone+92}

\section{The Brown Dwarf IMF from Turbulent Fragmentation}

In order to provide a quantitative estimate of the BD
abundance, a model for the structure of the density distribution is
required. There has been significant progress in the analytical
theory of supersonic turbulence in recent works
\citep{Boldyrev2002,Boldyrev+2002scaling,Boldyrev+2002structure}.
Some results regarding the
scaling of structure functions of the density field are already
available \citep{Boldyrev+2002structure,Padoan+2002scaling} and
could be used in the future for a rigorous analytical study of
the process of turbulent fragmentation.

A simple model of the expected mass distribution of dense cores
generated by supersonic turbulence has been proposed in
\cite{Padoan+Nordlund2002IMF}, on the basis of the two following assumptions:
i) The power spectrum of the turbulence is a power law; ii) the
typical size of a dense core scales as the thickness of the postshock
gas. The first assumption is a basic result for turbulent flows and
holds also in the supersonic regime
\citep{Boldyrev+2002scaling}.
The second assumption is suggested by the fact that postshock condensations
are assembled by the turbulent flow in a dynamical time.
Condensations of virtually any size can therefore be formed, 
independent of their Jeans' mass.

With these assumptions, together with the jump conditions for MHD
shocks (density contrast proportional to the Alfv\'{e}nic Mach
number of the shock), the mass distribution of dense cores
can be related to the power spectrum of turbulent velocity,
$E(k)\propto k^{-\beta}$. The result is the following
expression for the core mass distribution:
\begin{equation}
N(m)\,{\rm d}\ln m\propto m^{-3/(4-\beta)}{\rm d}\ln m ~.
\label{imf}
\end{equation}
If the turbulence spectral index $\beta$ is taken from the
analytical prediction \citep{Boldyrev+2002scaling}, which
is consistent with the
observed velocity dispersion-size Larson relation \citep{Larson79,Larson81}
and with our numerical results \citep{Boldyrev+2002scaling},
then $\beta \approx 1.74$ and the mass distribution is
\begin{equation}
N(m)\,{\rm d}\ln m\propto m^{-1.33}{\rm d}\ln m ~,
\label{salpeter}
\end{equation}
almost identical to the Salpeter stellar IMF \citep{Salpeter55}.
The exponent of the mass distribution is rather well constrained,
because the value of $\beta$ for supersonic turbulence cannot be
smaller than the incompressible value, $\beta= 1.67$ (sligthly larger
with intermittency corrections), and the Burgers case, $\beta=2.0$.
As a result, the exponent of the mass distribution is predicted to be 
well within the range of values of 1.3 and 1.5. In the following 
we use $\beta=1.74$, corresponding to a core mass distribution
$\propto m^{-1.36}$. 

While massive cores are usually larger than their critical mass,
$m_{\rm BE}$, the probability that small cores are dense enough to collapse is
determined by the statistical distribution of core density. In order to 
compute this collapse probability for small cores, we assume 
i) the distribution of core density can be approximated by the 
Log--Normal PDF of gas density and ii) the core density and mass are 
statistically independent. Because of the intermittent
nature of the Log-Normal PDF, even very small (sub--stellar) cores
have a finite chance to be dense enough to collapse.
Based on the first assumption, we can compute the distribution of the 
critical mass,  $p(m_{\rm BE})\,d m_{\rm BE}$, from the Log--Normal
PDF of gas density assuming constant temperature \citep{Padoan+97ext}.
The fraction of cores of  mass $m$ larger than their critical Bonnor--Ebert
mass is given by the integral of $p(m_{\rm BE})$ from 0 to
$m$. Using the second assumption of statistical independence of core density 
and mass, the mass distribution of collapsing cores is
\begin{equation}
N(m)\, {\rm d}\ln m\propto m^{-3/(4-\beta)}\left[\int_0^m{p(m_{\rm BE}){\rm d}m_{\rm BE}}\right]\,{\rm d}\ln m ~.
\label{imfpdf}
\end{equation}
The mass distribution is found to be a power law, determined by
the power spectrum of turbulence, for masses larger than approximately
1 M$_{\odot}$ (using physical parameters typical of molecular clouds).  
At smaller masses the mass distribution flattens, reaches a  maximum at 
a fraction of a solar mass, and then decreases with decreasing stellar mass. 

The upper panel of Figure~\ref{fig2} shows five mass distributions computed
from equation (\ref{imfpdf}).
Three of them (solid lines) are computed for $\langle n\rangle=10^4$~cm$^{-3}$, 
$T=10$~K and for three values of the sonic rms Mach number, 
$M_{\rm S}=5$, 10 and 20. An increase in the rms Mach number by a factor of two,
from $M_{\rm S}=5$ to $M_{\rm S}=10$, results in a growth of the abundance 
of 0.07~M$_{\odot}$ stars by more than a factor of ten (relative to stars of 
approximately 1~M$_{\odot}$ or larger). From $M_{\rm S}=10$ to $M_{\rm S}=20$,
the abundance of 0.07~M$_{\odot}$ stars increases by approximately a factor of
three. The other two mass distributions (dotted lines) are computed for 
$M_{\rm S}=10$, $T=10$~K and density 
$\langle n\rangle=5\times 10^3$~cm$^{-3}$ (lower plot), and 
$\langle n\rangle=2\times 10^4$~cm$^{-3}$ (upper plot). 

The IMF of the cluster IC 348 in Perseus, obtained by \cite{Luhman+2003}, is 
plotted in the lower panel of Figure~\ref{fig2} (solid line histogram). 
The IMF of this cluster has been chosen for the comparison with the theoretical 
model because it is probably the most reliable observational IMF including both 
brown dwarfs and hydrogen burning stars. Spectroscopy has been obtained for every 
star and the sample is unbiased in mass and nearly complete down to 0.03~$M_{\odot}$.
In the lower panel of Figure~\ref{fig2} we have also plotted the theoretical
mass distribution computed for $\langle n\rangle=5\times 10^4$~cm$^{-3}$ , 
$T=10$~K and $M_{\rm S}=7$. As discussed above, these parameters are appropriate
for the central $5\times 5$~arcmin of the cluster ($0.35\times 0.35$~pc),
where the stellar density corresponds to approximately $2\times 10^4$~cm$^{-3}$.
The figure shows that the theoretical distribution of collapsing 
cores, computed with parameters inferred from the observational data, 
is roughly consistent with the observed stellar IMF in the cluster IC 348. 

Similar IMFs were obtained for the Trapezium cluster in Orion by 
\cite{Luhman+2000} and for the inner region of the Orion Nebula Cluster 
by \cite{Hillenbrand+Carpenter2000}, using D'Antona and Mazzitelli's 
1997--evolutionary models. However, based on \cite{Baraffe+98} evolutionary 
models, these two IMFs contain a slightly larger abundance of brown dwarfs
than found in IC 348 and predicted by the theoretical model (unless larger
values of density or Mach number are assumed).
A larger BD abundance is found in $\sigma$ Orionis by
\cite{Bejar+2001}, while the IMFs obtained by \cite{Najita+2000} for IC348 
and the Pleiades' IMF \citep{Bouvier+98} are consistent with the IMF in the 
Orion Nebula Cluster. Several other IMFs of young clusters, including both
stellar and sub--stellar masses, have been recently 
obtained. 

The present theoretical model may in some cases underestimate the BD 
abundance, if a significant fraction of BDs are formed as members of binary 
systems, because the process of binary formation is not taken into account.
As an example, if most prestellar cores assembled by the turbulence 
were able to fragment into binary stars due to processes unrelated to turbulent
fragmentation, the final BD abundance would be increased, while
at larger masses the mass distribution would be indistinguishable from the 
one predicted by the model. 

\cite{Luhman2000} found that the number of BDs in Taurus is 12.8 times
lower than in the Trapezium cluster \citep{Luhman+2000}. This
result was based on a single BD detection and on several
low mass stars. The deficit of BDs in Taurus relative to the 
Trapezium cluster has been confirmed in a more recent work by
\cite{Briceno+2002}, although reduced to approximately a factor
of two between the BD abundance of Orion and Taurus.
The smaller relative abundance of BDs in Taurus may be explained 
by the analytical model as due to a decrease in the turbulent velocity 
dispersion (rms Mach number) or in the average gas density by less 
than a factor of two. This is consistent with the lower velocity 
dispersion and density in Taurus relative to Orion.

\section{Numerical Results} 

The mass distribution of prestellar condensations can be measured directly
in numerical simulations of supersonic turbulence. With a mesh of
250$^3$ computational cells, and assuming a size of the simulated
region of a few pc, it is not possible to follow numerically the
gravitational collapse of individual protostellar condensations. However,
dense cores at the verge of collapse can be selected in numerical
simulations by an appropriate clumpfind algorithm. We use an algorithm that selects
cores by scanning the full range of density levels. It eliminates large
cores that are fragmented into smaller and denser ones. Cores are also
excluded if their gravitational energy is not large enough to overcome
thermal and magnetic support against the collapse, because only collapsing
cores are selected.

A mass distribution of collapsing cores, derived from the density
distribution in a numerical simulation is shown in Figure~\ref{fig3}.
The mass distribution is computed from two snapshots of
a 250$^3$ simulation with rms Mach number $M_{\rm S}\approx 10$. 
We have used a random
external force on large scale and an isothermal equation of state (for
details of the numerical method see \cite{Padoan+Nordlund99MHD} and
references therein). The average gas density has been scaled to
500~cm$^{-3}$ and the size of the computational box to 10~pc. These
values have been chosen to be able to select condensations in a range of masses
from a sub--stellar mass to approximately 10~M$_{\odot}$.  With this
particular values of average gas density, size and resolution of the computational
box, the smallest mass that can be achieved numerically is
0.057~M$_{\odot}$. 

The analytical mass
distribution, $N(m)$, computed with the same physical parameters used in the
numerical simulation ($\langle n\rangle=500$~cm$^{-3}$, $T=10$~K and 
$M_{\rm S}=10$) is plotted in Figure~\ref{fig3} as a dashed line.
There is no free parameter to adjust the shape of the analytical function
and its mass scale, once the values of density, temperature, and rms
Mach number have been specified to agree with those assumed in the numerical
experiment. The agreement between the numerical and the analytical mass
distributions provides strong support for our simple analytical model of
the mass distribution of collapsing cores generated by supersonic
turbulence.

\section{Summary and Conclusions}    
 
In this work we have investigated the possibility that gravitationally 
unstable protostellar cores of BD mass are assembled by turbulent shocks. 
We have found that a fraction of BD mass cores formed by the turbulence 
are dense enough to collapse. The predicted BD abundance is consistent 
with the abundance observed in young stellar clusters if the theoretical 
IMF is computed with average density and rms sonic Mach number 
appropriate for dense cluster--forming regions inside molecular cloud 
complexes. 

We have not studied the evolution of turbulent density fluctuations
smaller than their critical mass. If subcritical fluctuations of 
BD mass are inside a larger collapsing core, they would be
increasing their density as the background collapses.  Additional 
fluctuations may also be created by turbulence during the collapse.
Under appropriate conditions, a fraction of these fluctuations may be 
able to collapse into additional BDs or giant planets. 

Future numerical simulations designed to study this process will
require not only a very large dynamical range of scales, possibly achieved only
by particle or adaptive mesh refinement codes, but also an accurate 
physical description of the supersonic turbulence including magnetic
forces.

\acknowledgements 

We are grateful to Kevin Luhman and Gilles Chabrier for valuable discussions on the
stellar IMF in clusters and to Bo Reipurth for pointing out a numerical
error in the definition of the critical mass. 
The work of {\AA}N was supported by a grant from the Danish
Natural Science Research Council. Computing resources were provided by the Danish
Center for Scientific Computing. 

\clearpage


\clearpage

\onecolumn

{\bf Figure captions:} \\

\begin{figure}
\centerline{\epsfxsize=13cm \epsfbox{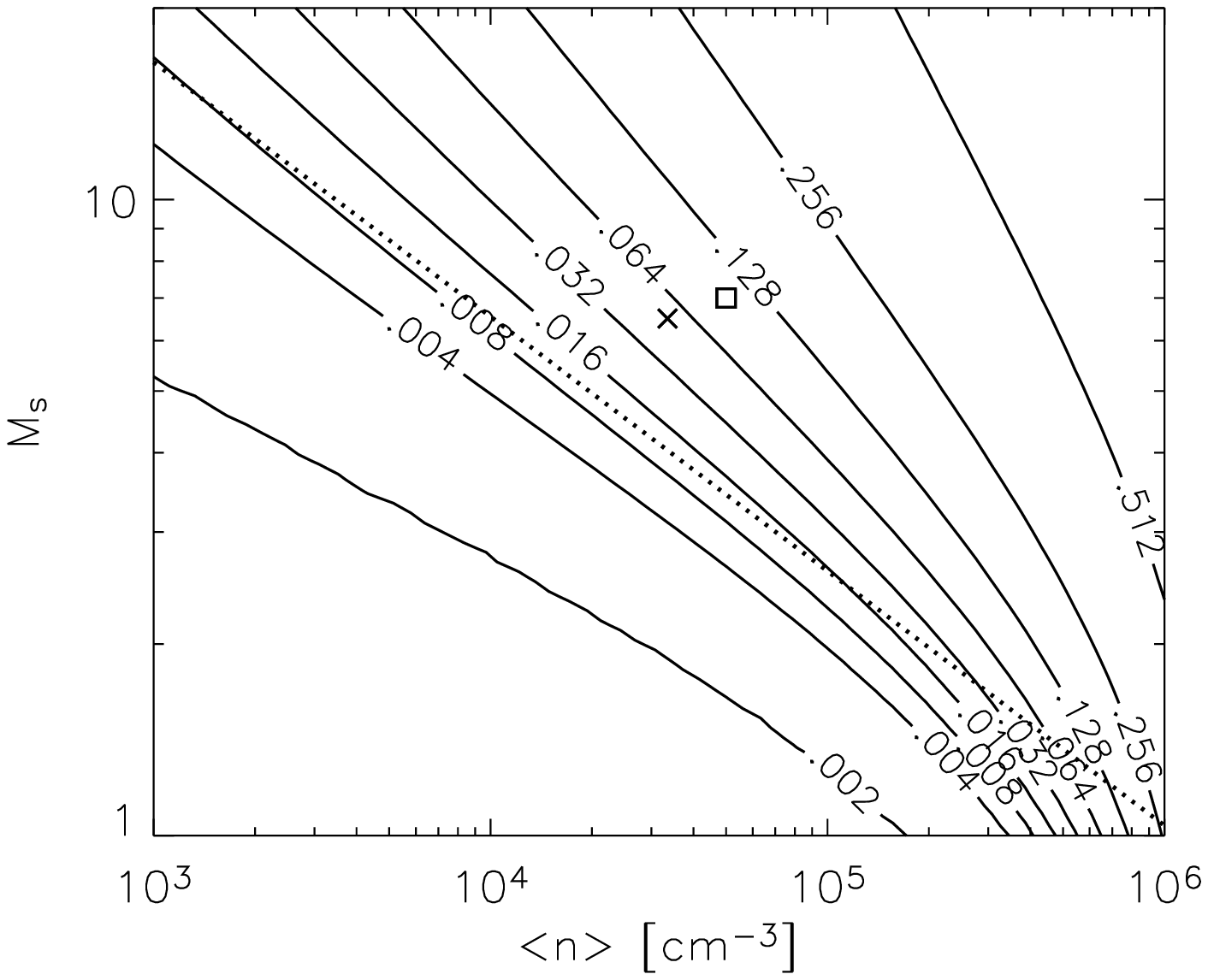}}
\caption[]{}
\label{fig1}
\end{figure}

{\bf Figure \ref{fig1}:} Contour plot of the fractional mass available
for the formation of BDs, $f_{\rm BD}$, on the plane of gas density--rms Mach
number, $\langle n \rangle$--$M_{\rm S}$. The dotted line corresponds to values
of rms Mach number and average gas density of typical Larson relations. 
The cross corresponds to the initial conditions in the simulation by 
\cite{Bate+2002} and the square to the physical parameters appropriate
for the central region of the young cluster IC 348 \citep{Luhman+2003}.\\

\begin{figure}
\epsfxsize=13cm \epsfbox{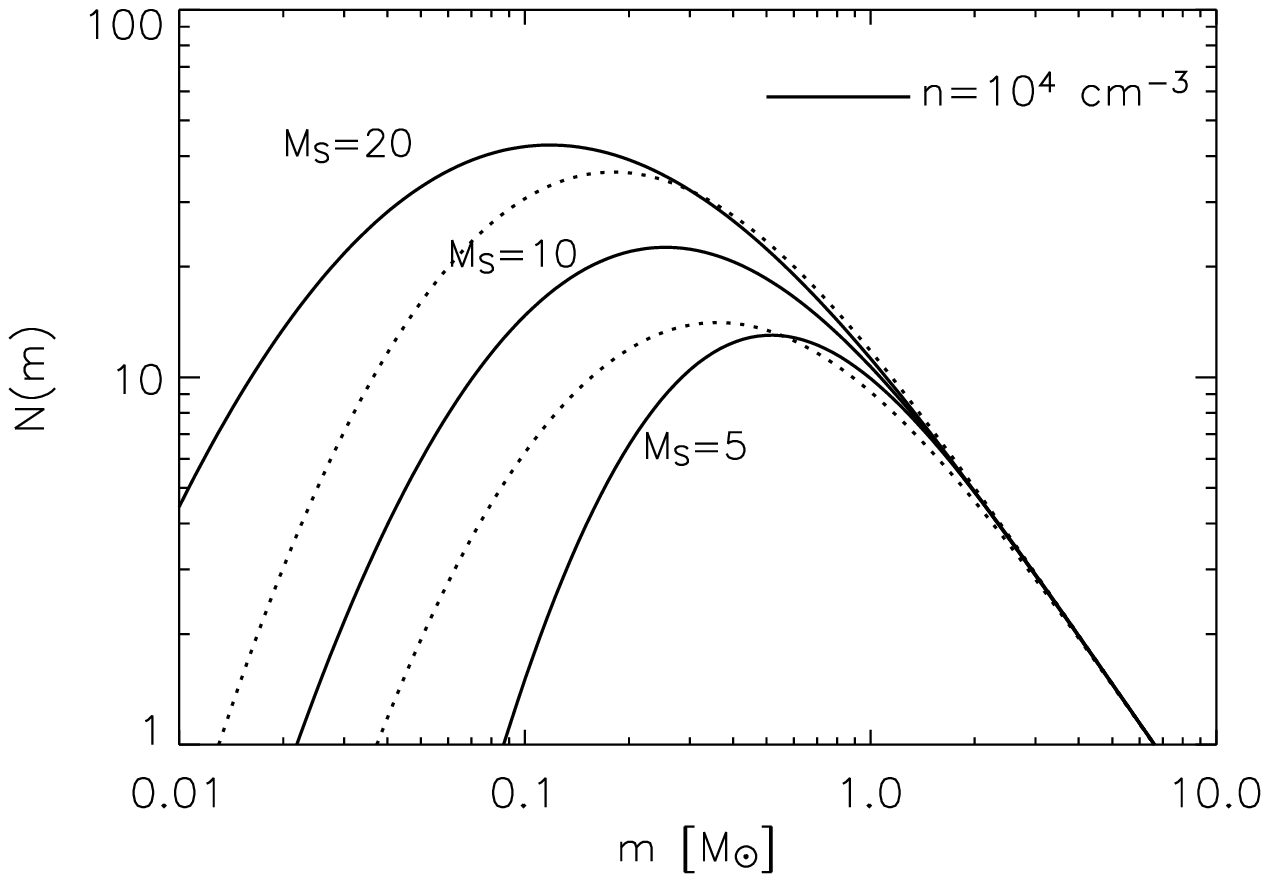}
\epsfxsize=13cm \epsfbox{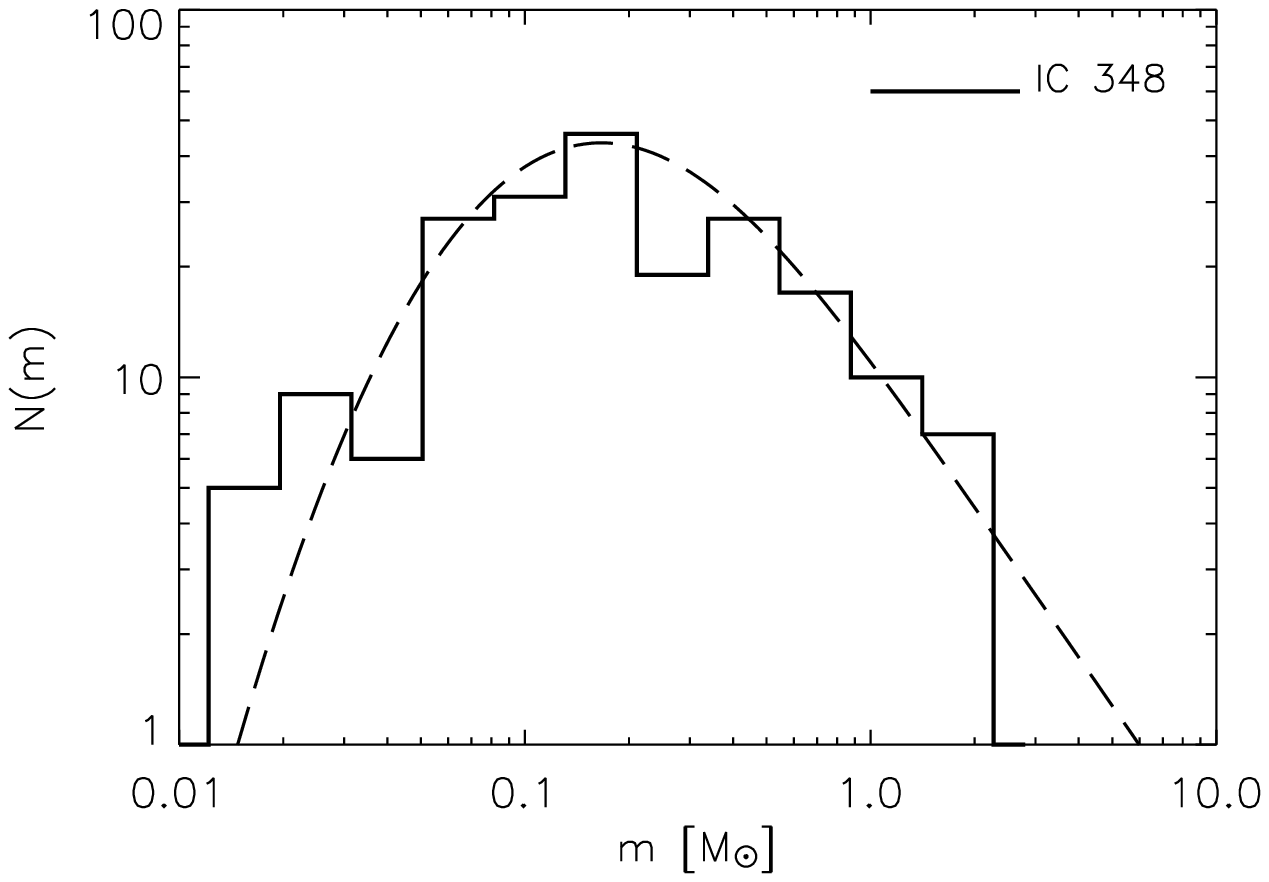}
\caption[]{}
\label{fig2}
\end{figure}

{\bf Figure \ref{fig2}:} Upper panel: Analytical mass distributions computed
for $\langle n\rangle=10^4$~cm$^{-3}$, $T=10$~K and 
for three values of the sonic rms Mach number, $M_{\rm S}=5$, 10 and 20
(solid lines). The dotted lines show the mass distribution for 
$T=10$~K, $M_{\rm S}=10$ and $\langle n\rangle=5\times10^3$~cm$^{-3}$
(lower plot) and $\langle n\rangle=2\times10^4$~cm$^{-3}$ (upper plot).
Lower panel: IMF of the cluster IC 348 in Perseus obtained by \cite{Luhman+2003} 
(solid line histogram) and theoretical IMF computed for 
$\langle n\rangle=5\times 10^4$~cm$^{-3}$, $T=10$~K and 
$M_{\rm S}=7$ (dashed line). The histogram of IC 348 mass function
in \cite{Luhman+2003} is computed with 9 bins, while the histogram shown 
here is computed with 12 bins. \\

\begin{figure}
\centerline{\epsfxsize=13cm \epsfbox{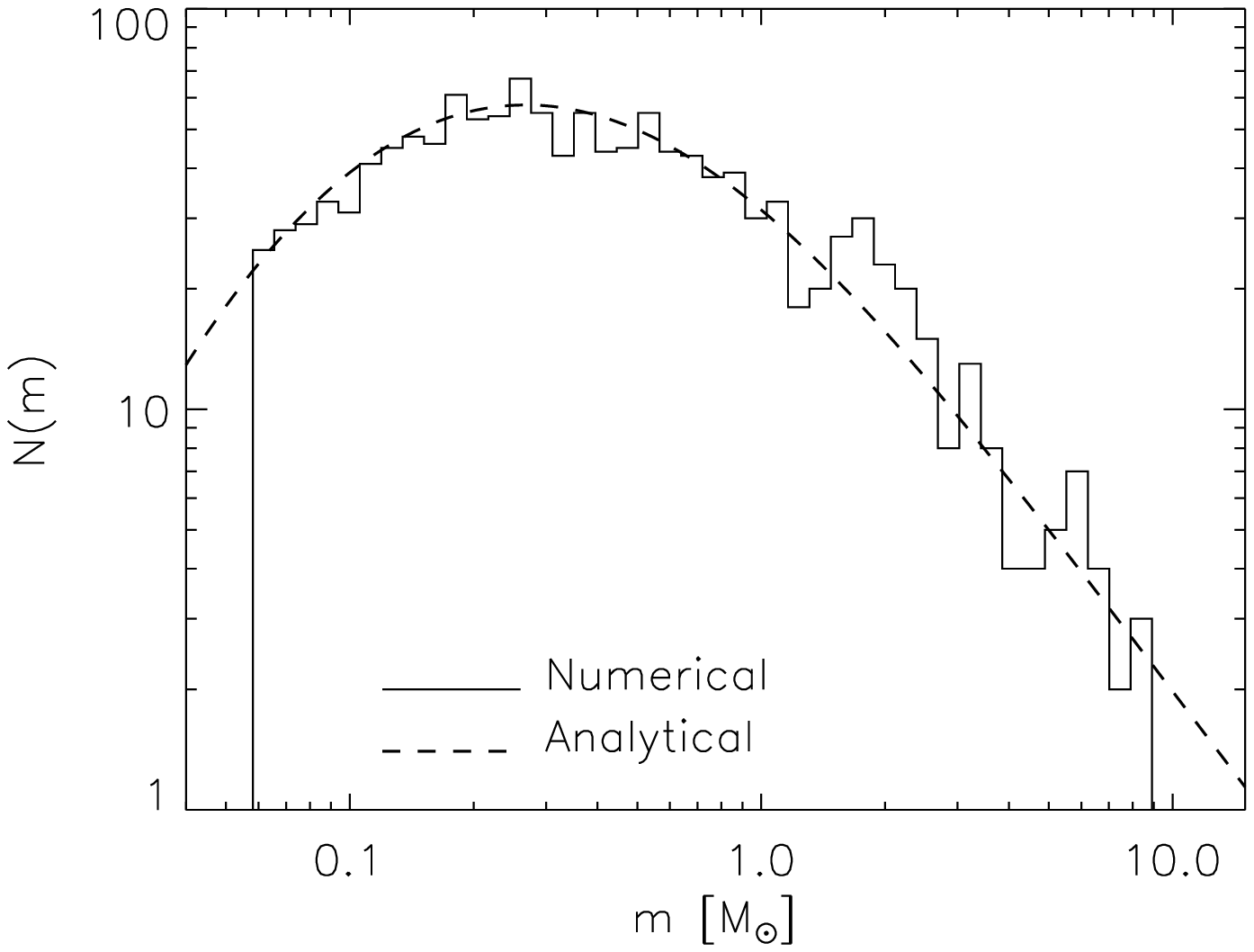}}
\caption[]{}
\label{fig3}
\end{figure}

{\bf Figure \ref{fig3}:} Solid line: Mass distribution of collapsing cores,
derived from the density distribution of two snapshots of
a 250$^3$ simulation with rms Mach number $M_{\rm S}\approx 10$,
external random forcing on large scale and isothermal equation
of state. The simulation is scaled to physical units assuming
$\langle n\rangle=500$~cm$^{-3}$, $T=10$~K, and a mesh size of 10~pc. 
The fractional mass in collapsing cores is 5\% of the total mass. 
Dashed line: Analytical mass distribution computed for $\langle n\rangle=500$~cm$^{-3}$,
$T=10$~K and $M_{\rm S}=10$. \\

\clearpage

\clearpage

\clearpage

\end{document}